\documentclass[aps,showkeys,preprint,prb,12pt]{revtex4}
\usepackage{epsfig}
\usepackage{graphicx}
\usepackage{amsmath,amssymb,amsfonts,latexsym}
\usepackage{graphicx}

\usepackage{epsfig,amssymb,amsfonts,verbatim}
\usepackage{amsmath}
\usepackage{latexsym}
\usepackage{amsfonts}
\usepackage{amssymb}
\usepackage{color}
\def\bfl{\begin{flushleft}}
\def\efl{\end{flushleft}}
\def\bfr{\begin{flushright}}
\def\efr{\end{flushright}}
\def\bc{\begin{center}}
\def\ec{\end{center}}

\def\ba{\begin{eqnarray}}
\def\ea{\end{eqnarray}}
\def\baa#1{\begin{array}{#1}}
\def\eaa{\end{array}}
\def\bw{\begin{widetext}}
\def\ew{\end{widetext}}

\def\text#1{\mbox{#1}}

\def\square{{\hbox{$\sqcup$}\llap{\hbox{$\sqcap$}}}}

\Large
\begin{document}

\title{Chemical Reaction due to Stronger Ramachandran Interaction}
\author{Andrew Das Arulsamy}
\affiliation{Condensed Matter Group, Division of Interdisciplinary Science, F-02-08 Ketumbar Hill, Jalan Ketumbar, 56100 Kuala-Lumpur, Malaysia}


\date{\today}

\begin{abstract}
The origin of a chemical reaction between two reactant atoms is associated to the activation energy, with the assumption that, high-energy collisions between these atoms, are the ones that overcome the activation energy. Here, we (i) show that a stronger attractive van der Waals (vdW) and electron-ion Coulomb interactions between two polarized atoms are responsible to initiate a chemical reaction, either before or after the collision. We derive this stronger vdW attraction formula exactly using the quasi one-dimensional Drude model within the ionization energy theory and the energy-level spacing renormalization group method. Along the way, we (ii) expose the precise physical mechanism responsible for the existence of a stronger vdW interaction for both long and short distances, and also show how to technically avoid the electron-electron Coulomb repulsion between polarized electrons from these two reactant atoms. Finally, we properly and correctly associate the existence of this stronger attraction to Ramachandran's `normal limits' (distance shorter than what is allowed by the standard vdW bond) between chemically nonbonded atoms.
~~\\
\textbf{Running title}: Ramachandran Interaction Between Atoms ~~\\
\textbf{email}: sadwerdna@gmail.com
\end{abstract}

\maketitle

\section*{1. Introduction}

Chemical reaction is one of the most remarkable quantum mechanical phenomena that made biological life possible on earth, which can also be manipulated to discover novel `green' chemistry, and to avoid `toxic' technologies, provided we can understand the reaction microscopically to a certain extent. However, within the current formalism of chemical kinetics and the laws of thermodynamics, a chemical reaction between two reactant atoms (for a given chemical potential, temperature, pressure and concentration) occurs when they collide with each other with enough energy to overcome an Arrhenius-type activation energy, to form a(several) new bond(s)~\cite{iupac,selva,senthil}. Here, we attempt to go beyond this macroscopic collision-thermodynamic based description by incorporating some microscopic physics (with respect to atomic energy level spacings) to understand why and how a chemical reaction proceeds for a given set of conditions.

The chemical reaction considered here may need the right reaction coordinates to form the transition state~\cite{tiwar}. Therefore, for two atoms (either isolated or from two different molecules) to chemically react, they need to have the correct reaction coordinates, and even after fulfilling this condition, these atoms still need to have an overall strong attractive interaction (either before or after the collision) to form the required transition state. For example, the transition state is by definition gives equal probability for the reaction to go either way (to form reactant atoms, or the reaction product). Therefore, there should be a microscopic mechanism that is responsible for an effective attraction between reactant atoms such that the reaction product is more likely than the formation of reactant atoms, even after forming the transition state. The standard attractive vdW interaction is a long-range type, and it is too weak (because it is proportional to 1/$R^6$) to initiate any chemical reaction. On the other end, we have the electron(from atom 1)-ion(from atom 2) Coulomb attraction, which is the strongest, but cannot lead to an attraction due to electron(from atom 1)-electron(from atom 2) Coulomb repulsion. Apparently, collisions among the reactant atoms can overcome this electron-electron Coulomb repulsion if the atoms are weakly interacting. Collisions can induce electronic excitation and/or polarization, which then can activate (i) the electron-electron Coulomb repulsion or (ii) the attraction due to ion-ion vdW and electron-ion Coulomb interactions. As a consequence, we need to find the microscopic physical origin of a chemical reaction with respect to the above strong attractions, which goes beyond the present-day thermodynamics (chemical potential) and energetic-collision knowledge. 

Here, we prove the existence of a stronger attraction (stronger than the usual vdW type) that can be used to understand why and how a chemical reaction can proceed (with further assistance from collisions) by overcoming the electron-electron Coulomb repulsion. In other words, we set out to prove the existence of a stronger vdW interaction, which is also shown to be further enhanced by the electron-ion Coulomb attraction to activate the chemical reaction. In particular, we expose that these stronger attractions are polarization-induced effect, which can be precisely associated to atomic energy-level spacings, and these stronger attractions are also responsible to overcoming the said repulsion (or the activation energy). Once the repulsion is overcome, either after, or before the collision, then it is straightforward to understand why and how a chemical reaction between two reactant atoms can proceed. We stress here that we do need atom-atom collisions for strongly interacting atoms (with isotropic polarization), and also for weakly polarized atoms. However, after the collision, the above attractions still need to be activated for a chemical reaction to proceed. 

It is to be noted here that this stronger vdW attraction has also been partly predicted computationally and indirectly~\cite{jcs1} by means of the Hermansson blue-shifting hydrogen bond~\cite{her1,her2,her3}, the Arunan composite hydrogen-vdW bond~\cite{arun,arun2,arun3,arun4} and the non-covalent carbon bond~\cite{arun5}. In the Section, Additional Notes prior to conclusions, we will show that Ramachandran, Sasisekharan and Ramakrishnan~\cite{ramc,ramc2,dicker} have already discovered the notion of stronger non-covalent bonds (stronger than the vdW bond) between atoms. This means that the Hermansson blue-shifting hydrogen bond, the hydrogen-vdW composite bond and the non-covalent carbon bond are special cases of the Ramachandran bond, in the absence of wavefunction overlapping or chemical reaction. Hence, the Ramachandran bond is a non-chemical bond that does not require chemical reaction.   

However, to pin down the origin of a stronger vdW and an attractive electron-ion Coulomb interactions that overcome the electron-electron repulsion (due to polarization), we definitely need analytic solutions and some abstract analyses. Therefore, we exploit the crude quasi one-dimensional (1D) Drude model within the ionization energy theory (IET)~\cite{pra} using the energy-level spacing renormalization group method~\cite{aop} to arrive at accurate and exact results. The `quasi' here implies that one can realign the vectors in a one-dimensional system to account for the two and three dimensional cases in which the atoms are treated as real atoms. In our further analyses (I and II), we unequivocally show why and how the stronger vdW interaction is also consistent with respect to the second-order perturbation theory and the $1/R$ multipole expansion. Therefore, its existence is not due to some interpretational issues arising exclusively from the crude Drude model Hamiltonian.

\section*{2. Theoretical details}

\subsubsection*{{\rm 2.1}~\textit{Drude model Hamiltonian}}

The relevant Hamiltonian for the one-dimensional two-atom system (strictly neutral atoms, but with polarizable valence electrons) plotted schematically in Fig.~\ref{fig:1} is given by~\cite{griffiths,stone} 
\begin {eqnarray}
&&H = H_{\rm O} + H_{\rm I}, \label{eq:1new} \nonumber \\&& H_{\rm O} = \frac{p_1^2}{2m} + \frac{kr^2_1}{2} + \frac{p_2^2}{2m} + \frac{kr^2_2}{2}, \nonumber \\&& 
H_{\rm I} = \frac{1}{4\pi\epsilon_0}\bigg[\frac{(e)(e)}{R} + \frac{(-e)(e)}{R + r_2} + \frac{(-e)(e)}{R - r_1} + \frac{(-e)(-e)}{R - r_1 + r_2}\bigg], \label{eq:2}
\end {eqnarray}  
where $p_{1,2}$ are the electrons momenta, and the ground state energy for the above Hamiltonian is (taking $R \gg r_{1,2}$ and after some nontrivial substitutions)~\cite{griffiths,stone} 
\begin {eqnarray}
E = \frac{1}{2}\hbar\bigg[\sqrt{\frac{k - (e^2/2\pi\epsilon_0R^3)}{m}} + \sqrt{\frac{k + (e^2/2\pi\epsilon_0R^3)}{m}}\bigg], \label{eq:3}
\end {eqnarray}  
where $m$ is the electron mass and $k$ is the interaction constant (that changes for different atoms). Note here that $\textbf{r}_1$ and $\textbf{r}_2$ are indeed vectors, but we have aligned the atoms such that their electronic polarization are maximum along a straight line where $\textbf{r}_1 \rightarrow r_1$ and $\textbf{r}_2 \rightarrow r_2$. We now prove why this alignment is valid for three-dimensional cases (imagine two polarizable spheres)---suppose $\textbf{r}_1$ (from sphere 1) is aligned along the $x$ axis, so that $|\textbf{r}_1| = r_1$ . The induced polarization due to $\textbf{r}_1$ causes $|\textbf{r}_2| = r_2(\theta,\phi)$ where $\theta < \pi/4$, $\phi < \pi/4$, $\theta$ is the angle in the $xy$-plane and $\phi$ denotes the angle in the $xz$-plane. It is easy to verify that if $\theta \geq \pi/4$ and $\phi \geq \pi/4$, then there is an additional induced polarization affecting $|\textbf{r}_2| = r_2(\theta,\phi)$. If such a case exists, then we just need to take this additional polarization into account (that affects $\textbf{r}_2$), and realign $\textbf{r}_2$ (from sphere 2) with respect to this new $\textbf{r}_1'$. In this second case, $|\textbf{r}_1'| = r_1'$ causes $|\textbf{r}_2'| = r_2'(\theta,\phi)$ where $\theta < \pi/4$ and $\phi < \pi/4$. This means that, $\theta$ and $\phi$ remains the same such that $\theta \in (0,\pi/4]$ and $\phi \in (0,\pi/4]$. For one-dimensional cases, $\theta = 0 = \phi$, while for a two-dimensional case, one has $\theta \in (0,\pi/4]$ and $\phi = 0$. $\square$

In Eq.~(\ref{eq:3}), $\hbar$ is the Planck constant divided by 2$\pi$, $R$ is the distance between two neutral and identical atoms. The subscripts 1 and 2 refer to electron 1 and 2 bounded to nuclei 1 and 2, respectively. The ordinary Hamiltonian denoted by $H_{\rm O}$ consists of two kinetic energy terms, and two semiclassical harmonic-oscillator type electron-ion potential terms, and they are associated to two electrons bounded to their respective nucleus (see Fig.~\ref{fig:1}). The interaction Hamiltonian, $H_{\rm I}$ captures all the Coulomb interaction between the two atoms depicted in Fig.~\ref{fig:1}. Note that $H_{\rm I}$ does not consider the Coulomb forces within the atom (between the valence electrons and their respective nucleus), which have been taken into account by $H_{\rm O}$. 

Now, one can renormalize the interaction potential constant, $k$ such that~\cite{aop}
\begin {eqnarray}
\tilde{k} = k\exp{[\lambda\xi]}, \label{eq:4}
\end {eqnarray}  
where $\lambda = (12\pi\epsilon_0/e^2)a_B$, in which, $a_B$ is the Bohr radius of atomic hydrogen, $e$ and $\epsilon_0$ are the electron charge and the permittivity of free space, respectively. The above renormalization procedure is also related (exactly) to the Shankar renormalization technique~\cite{shank1,shank2,shank3}. The renormalized interaction potential constant, $\tilde{k}$ (from Eq.~(\ref{eq:4})) replaces the standard $k$ in $H_{\rm O}$ to capture the changes to $k$ whenever one changes the type of atom. Whereas, the variable, $\xi$ denotes the atomic energy level spacing when $H_{\rm I} \neq 0$, which is proportional to $\xi$ when $H_{\rm I} = 0$. This proportionality (see Eq.~(\ref{eq:IN14b})) has been proven in Ref.~\cite{pra}, and we do not allow any electronic wavefunction overlapping.

If the Drude model Hamiltonian~\cite{stone,drude,mar,jones} is independent of $k$, then one should renormalize the frequency, such that~\cite{aop}, $\tilde{\omega} = \omega\exp{[\frac{1}{2}\lambda\xi]}$. Since the factor $\exp{[\frac{1}{2}\lambda\xi]}$ is a constant for a given system and for a given set of conditions, one will obtain the identical vdW interaction energy (see Eq.~(\ref{eq:6})) such that this energy is proportional to $(1/R^6)\exp{[-\frac{\texttt{s}}{2}\lambda\xi]}$, $\texttt{s} = 3$ regardless whether one uses the $k$-dependent Hamiltonian given by Eq.~(\ref{eq:2}) or the $\omega$-dependent Drude model Hamiltonian, provided that the condition, $k\exp{[\lambda\xi]} \gg e^2/2\pi\epsilon_0R^3$ is satisfied. If $k\exp{[\lambda\xi]} \gg e^2/2\pi\epsilon_0R^3$ is not satisfied, then one may obtain $\texttt{s} \neq 3$ where $\texttt{s}$ denotes a positive integer ($\texttt{s} > 0$). The proper renormalization procedure (with explanation) are given in Ref.~\cite{aop}.  

\subsubsection*{{\rm 2.2}~\textit{Ionization energy theory and its approximation}}

We now expose some details on the meaning of $\xi$ and the ionization energy approximation, which can be understood from the IET-Schr$\ddot{\rm o}$dinger equation~\cite{aop}, 
\begin {eqnarray}
&&i\hbar\frac{\partial \Psi(\textbf{r},t)}{\partial t} = \bigg[-\frac{\hbar^2}{2m}\nabla^2 + V_{\rm IET}\bigg]\Psi(\textbf{r},t) = H_{\rm IET}\Psi(\textbf{r},t) = (E_0 \pm \xi)\Psi(\textbf{r},t), \label{eq:IN14}
\end {eqnarray}  
where $\hbar = h/2\pi$, $h$ denotes the Planck constant and $m$ is the mass of electron, and $\Psi(\textbf{r},t)$ is the time-dependent many-body wavefunction. The eigenvalue, $E_0 \pm \xi$ is the real (true and unique) energy levels for a given quantum system (for a given molecule). Here, $E_0$ is a constant because it represents the energy level spacings at zero temperature, and in the absence of any external disturbances. On the other hand, $\xi = \xi^{\rm quantum}_{\rm matter}$ is the so called ionization energy. Therefore, all we need to do now is to find a way to obtain $\xi^{\rm quantum}_{\rm matter}$. 

We make use of the ionization energy approximation to approximate $\xi^{\rm quantum}_{\rm matter}$. This approximation hinges on this proportionality,
\begin {eqnarray}
&&\xi^{\rm quantum}_{\rm matter} \propto \xi^{\rm constituent}_{\rm atom}. \label{eq:IN14b}
\end {eqnarray}  
As stated earlier, $\xi^{\rm quantum}_{\rm matter}$ represents the real (unique and true) energy level spacings of a particular quantum system, and it can be regarded as a generalized electronic energy gap. In other words, $\xi^{\rm quantum}_{\rm matter}$ can only be obtained if we know the real many body wavefunction ($\Psi(\textbf{r},t)$), not some guessed wavefunctions, because only the true wavefunction can represent the electrons properly, and consequently, can capture the real electronic properties of a quantum matter. {\color{blue} Apart from that, note that the wave nature of the electron exists in Eq.~(\ref{eq:IN14b}) by definition---(i) $\xi$ denotes real (unique and true) values obtained from the real atomic spectra, and (ii) we have proven the existence of this correspondence~\cite{qpt}, $\Psi(\textbf{r},t) \rightarrow \xi(\textbf{r},t)$, which is also obvious from Eq.~(\ref{eq:IN14}). In particular, the ionization energy or the eigenvalue is a measure of the electron energy, and it depends on both the wave- and particle-nature of the electrons, regardless whether this energy is a measured or a calculated quantity using a wavefunction. In our case, the ionization energy is a measured quantity, and this does not imply the wave nature of the electron is not taken into account. 

In this work, we develop the relevant mechanism and construct the interaction potential term, which are responsible to induce chemical reaction between two atoms even in the absence of collisions. Therefore, one cannot list the pros and cons for a proposed mechanism or for a constructed proof because they (the pros and cons) do not exist. However, one can write down the advantages and disadvantages of a theory. For example, if the theory is applied with some guessed functions, then we do have the pros and cons because calculating a number by this method does not expose the generalized mechanism of a process, and it is system-dependent. This means that, one can always go back and change the functions here and there to calculate a new number that compare favorably with experimental data. Such changes are not generalizable, and therefore, they are system dependent and consequently, we need to define the pros and cons with respect to these changes. For example, these changes are suitable for system A, those changes are for system B, and so on. On the contrary, our work is entirely based on the first principles with well-defined operators, approximation and analytic method, and therefore, we never rely on any guessed functions. The restriction here is that the chemical reaction cannot occur if the above mechanism or proof is not satisfied, with or without collisions. The advantages and disadvantages of the theory employed here, namely, IET, including a fully worked out example are given elsewhere~\cite{jcs1}. The example in Ref.~\cite{jcs1} considered the interactions between different atoms, namely, between H and O, C and N, and between O and P. The responsible atoms interact maximally to induce chemical reactions (based on the principle of maximum interaction~\cite{jcs1}), provided that certain conditions are met (see the introduction). Apart from that, the stronger Ramachandran attraction has also been applied in cation channels, which reproduced the Eisenman sequence exactly, including the generalized mechanisms of ion selectivity~\cite{ionchan}. Other examples in solid state chemistry have been reported in Refs.~\cite{radha,dt,jap,pccc,kiran,sree,balaji}.} 

In molecules, $\xi^{\rm quantum}_{\rm matter} = \xi_{\rm molecule}$ is the energy level spacing between an occupied level (in the highest occupied molecular orbital (HOMO)) and an empty level (in the lowest unoccupied molecular orbital (LUMO)). Using this approximation, we can now predict the changes to $\xi^{\rm quantum}_{\rm matter}$ if we know $\xi^{\rm constituent}_{\rm atom}$. The values for $\xi^{\rm constituent}_{\rm atom}$ can be readily obtained from the experimental atomic spectra. The ionization energy approximation becomes exact if the quantum matter are atoms or ions ($\xi^{\rm quantum}_{\rm matter} = \xi^{\rm atom}_{\rm ion}$ from Eq.~(\ref{eq:IN14b})). The approximation given in Eq.~(\ref{eq:IN14b}) has been proven---the indirect proof reads, $H_{\rm IET}\Psi(\textbf{r},t) = (E_0 \pm \xi^{\rm quantum}_{\rm matter})\Psi(\textbf{r},t) \propto (E_0 \pm \xi^{\rm constituent}_{\rm atom})\Psi(\textbf{r},t)$, while the second formal proof is direct based on logic and the excitation probability of electrons and holes within the ionization energy based Fermi-Dirac statistics~\cite{aop}. 

\subsubsection*{{\rm 2.3}~\textit{Renormalized vdW interaction}}

From here onwards, any variable found to wear a tilde implies a renormalized parameter. Subsequently, $E$ (from Eq.~(\ref{eq:3})) can be written in the form (after making use of the series, $\sqrt{1 + x} = 1 + x/2 - x^2/8 + x^3/16 - \ldots$),
\begin {eqnarray}
\tilde{E}(\xi) &=& \frac{\hbar}{2}\bigg[\bigg(1 + \bigg(-\frac{e^2}{2\pi\epsilon_0R^3}\bigg)\frac{1}{2\tilde{k}} - \bigg(-\frac{e^2}{2\pi\epsilon_0R^3}\bigg)^2\frac{1}{8\tilde{k}^2} + \ldots\bigg) \nonumber \\&& + \bigg(1 + \bigg(\frac{e^2}{2\pi\epsilon_0R^3}\bigg)\frac{1}{2\tilde{k}} - \bigg(\frac{e^2}{2\pi\epsilon_0R^3}\bigg)^2\frac{1}{8\tilde{k}^2} + \ldots \bigg)\bigg]\sqrt{\frac{\tilde{k}}{m}}, \label{eq:5}
\end {eqnarray}  
in which, we consider only the first three terms by imposing the condition, $k\exp[\lambda\xi] \gg e^2/2\pi\epsilon_0R^3$, in other words $(e^2/2\pi\epsilon_0R^3)^{\texttt{n}}\tilde{k}^{-\texttt{n}} \approx 0$ for $\texttt{n} \geq 4$ and $\texttt{n} \in \mathbb{N}_{\rm even}$ where $\mathbb{N}_{\rm even}$ is the set of even natural numbers, and finally, Eq.~(\ref{eq:5}) also satisfies $|\pm e^2/2\pi\epsilon_0R^3\tilde{k}| \leq 1$. Using $\tilde{\omega}_0 = \sqrt{\tilde{k}/m}$ we can carry out the subsequent algebraic rearrangements of Eq.~(\ref{eq:5}). After subtracting $\hbar\tilde{\omega}_0$, Eq.~(\ref{eq:5}) will lead us to 
\begin {eqnarray}
&&\tilde{E}(\xi) - \hbar\tilde{\omega}_0 = \tilde{V}^{\rm std}_{\rm Waals}(\xi) = \bigg\{-\frac{\hbar}{8m^2\omega_0^3}\bigg(\frac{e^2}{2\pi\epsilon_0}\bigg)^2\frac{1}{R^6}\bigg\}\exp\bigg[-\frac{3}{2}\lambda\xi\bigg]. \label{eq:6}
\end {eqnarray}  
This is the standard vdW interaction energy in its renormalized form (due to the renormalizing factor (renormalizer for short), $\exp[-(3/2)\lambda\xi]$), and note the popular vdW factor, $1/R^6$. The original (unrenormalized) vdW formula is given in the curly bracket. However, even in the presence of this renormalizer, Eq.~(\ref{eq:6}) is only valid for weakly interacting systems because of this condition $k\exp[\lambda\xi] \gg e^2/2\pi\epsilon_0R^3$. We will explain why this is so in the following section. 

The term, $\hbar\tilde{\omega}_0$ is the renormalized ground state energy ($\tilde{E}(\xi)$) when $H_{\rm I} = 0$ (in the absence of Coulomb interactions between charges from different atoms or molecules). One can observe the correct trend from Eq.~(\ref{eq:6}), in which, for a given $R$, when $\xi$ shoots to infinity (valence electrons with zero polarization), $\tilde{V}^{\rm std}_{\rm Waals}(\xi) \rightarrow 0$, as it should be. In contrast, $\tilde{V}^{\rm std}_{\rm Waals}(\xi) \rightarrow$ maximum, if $\xi$ obtains an allowable minimum value. Note here that taking $\xi = 0$ is physically not acceptable as this means that the valence electrons are not bound to their respective nucleus, and they are free. 
\begin{figure}
\begin{center}
\scalebox{0.4}{\includegraphics{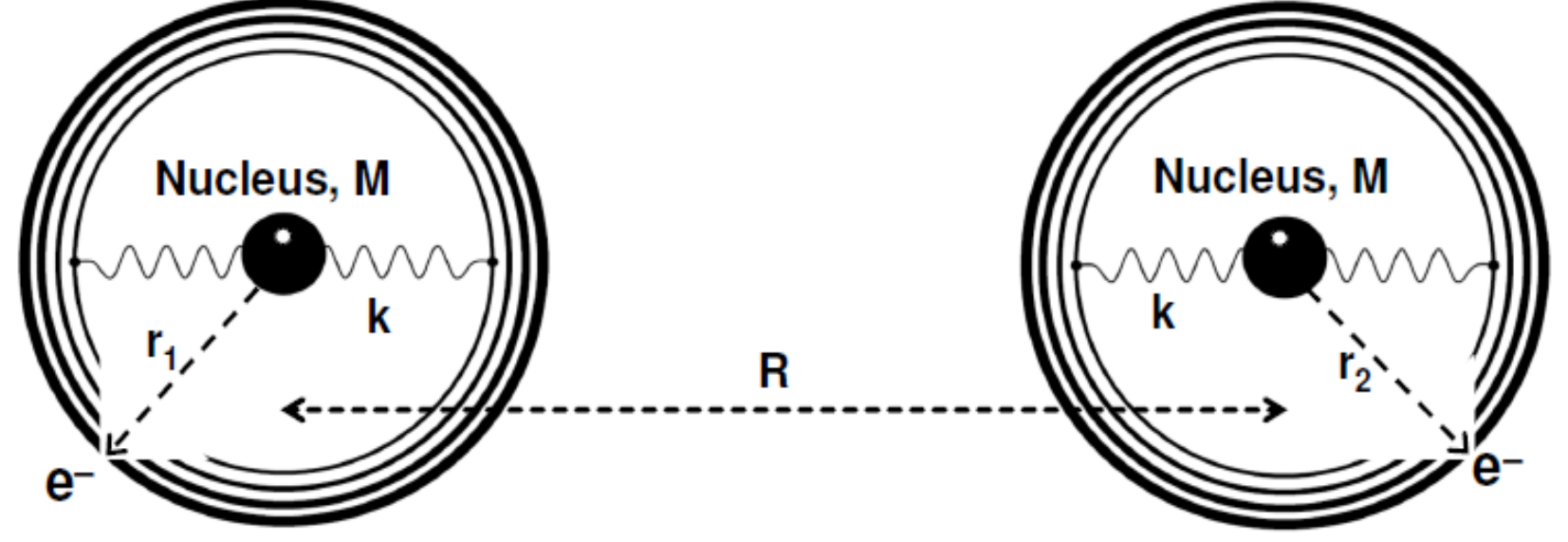}}
\caption{Two identical atomic system with discrete energy levels are sketched where $r_{1,2}$, $M$ and $k$ denote the coordinate for the valence electrons, $e^-$, the ionic mass and the spring (intermolecular interaction potential) constant, respectively. The discrete and quantized energy levels strictly imply non-free electrons. Here, $R$ is the distance between those two nuclei, and $M \gg m$ where $m$ is the mass of a valence electron. We set the coordinate such that $r_{1,2}$ and $R$ are all on the $x$-axis where the charges are arranged in a straight horizontal line, in the following order, $+ ~- ~+ ~-$. Negative charges correspond to the valence electrons from the left and right atoms, while the positive charges are from the respective nucleus. In the text, we also denote the left atom as atom 1, whereas atom 2 is to the right.}
\label{fig:1}
\end{center}
\end{figure}

\subsubsection*{{\rm 2.4}~\textit{Stronger vdW interaction formula}}

\begin{figure}
\begin{center}
\scalebox{0.46}{\includegraphics{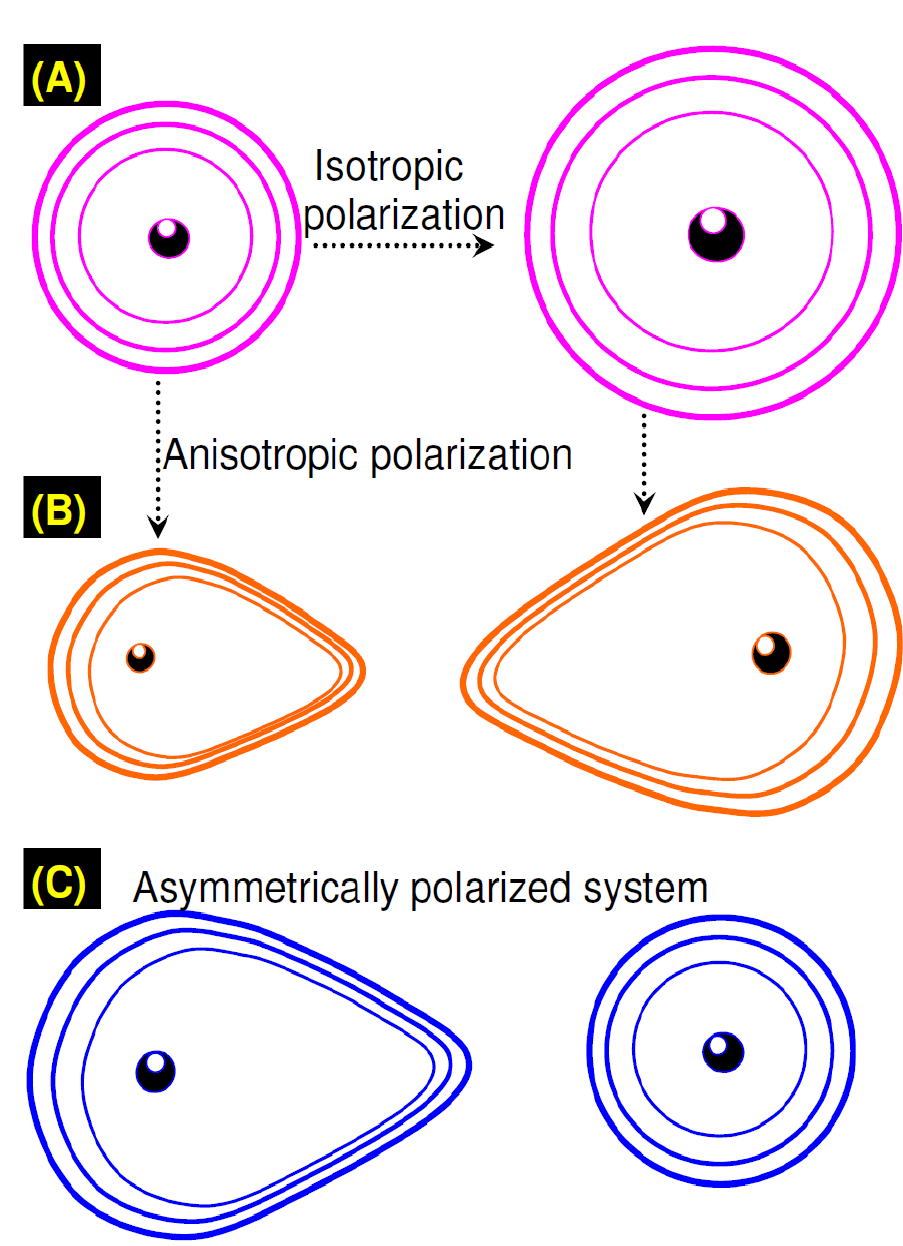}}
\caption{We define here the different types of polarizations discussed in the text. The sizes and shapes of the sketches are not to scale. The nuclei are at the center, while the lines represent the energy-levels, similar to Fig.~\ref{fig:1}. (A) The atom on the left is prior to polarization. After intense laser or at a higher temperature, the atom is polarized isotropically as we have sketched on the right-hand side. The two atoms sketched in (A) can also be anisotropically polarized as sketched in (B), if and only if these two atoms interact strongly. Therefore, anisotropic polarization is an induced polarization. This is often the case for strongly interacting systems such that there exists an effective electron-electron Coulomb repulsion leading to a repulsion between these atoms. Apart from that, (C) asymmetric polarization can also occur if one of the atom is highly polarizable, while the other atom is the least polarizable one, which will give rise to an effective attractive interaction. Again, this attraction is only valid for strongly interacting systems. However, if these two atoms interact weakly, then they are approximately isotropically polarized (regardless whether the polarization is large or small), giving rise to the standard vdW attractive interaction for $R > r_1 + r_2$. Here, $R$ is the internuclear distance, while $r_1$ and $r_2$ denote the radii of the two isotropically polarized atoms. Here, Fig.~\ref{fig:2} contains two-dimensional diagrams for clarity, but their analyses apply regardless whether a given system is one-, two- or three-dimensional.}
\label{fig:2}
\end{center}
\end{figure}

We now prove the existence of a stronger attraction between two interacting atoms. The above standard renormalized vdW attraction (see Eq.~(\ref{eq:6})) is weak because $\tilde{V}^{\rm std}_{\rm Waals}(\xi) \propto 1/R^6$ such that the polarization is either small or, if the polarization is large, then it is approximately isotropic (see Fig.~\ref{fig:2}(A)) because the atoms are weakly interacting, made sure by the condition, $k\exp{[\lambda\xi]} \gg e^2/2\pi\epsilon_0R^3$ or $k \gg e^2/2\pi\epsilon_0R^3$. In the case of strongly interacting two-atom system, a larger induced anisotropic polarization is obtainable (see Fig.~\ref{fig:2}(B)). Here, Fig.~\ref{fig:2}(A) depicts a weakly interacting system with an isotropically polarized atom, while Fig.~\ref{fig:2}(B) and (C) depict two types of strongly interacting systems with anisotropic polarization. In particular, (A) depicts isotropic polarization that is valid for weakly interacting atoms, (B) is for strongly interacting atoms, which induce an anisotropic polarization and also an electron-electron Coulomb repulsion between atoms, while (C) contains a mixture of easily-polarizable and least-polarizable atoms, giving rise to an asymmetric polarization. Here, (A) and (B) need additional assistance to initiate the strong attractions, namely, collisions between reactant atoms. In contrast, (C) may not need any external `help', hence could be spontaneous. To induce such an anisotropic polarization, we need a stronger interaction between these atoms such that the condition, $R \gg r$ is not violated (to make sure Eq.~(\ref{eq:3}) is still valid). In other words, we need a new condition, $k\exp{[\lambda\xi]} = e^2/2\pi\epsilon_0R^3$, not $k\exp{[\lambda\xi]} \gg e^2/2\pi\epsilon_0R^3$ to impose a large anisotropic polarization. The new condition allows a large intermolecular interaction energy if $\xi$ is small (see Eq.~(\ref{eq:6})). The only analytic relation with the smallest $\xi$ is coincidently $k\exp{[\lambda\xi]} = e^2/2\pi\epsilon_0R^3$ where $k\exp{[\lambda\xi]} < e^2/2\pi\epsilon_0R^3$ is physically not allowed. 

After making use of this new condition, $k\exp{[\lambda\xi]} = e^2/2\pi\epsilon_0R^3$, we can rewrite Eq.~(\ref{eq:3}) to obtain (subject to the renormalization procedure developed in Ref.~\cite{aop})
\begin {eqnarray}
&&\tilde{E}(\xi) - \hbar\tilde{\omega}_0 = \tilde{V}_{\rm Waals}(\xi) = \frac{1}{2}\hbar\bigg[\sqrt{\frac{2\tilde{k}}{m}}\bigg] - \hbar\tilde{\omega}_0. \label{eq:8}
\end {eqnarray}  
Unlike Eq.~(\ref{eq:6}), which was approximated using Eq.~(\ref{eq:5}), Eq.~(\ref{eq:8}) is obtained exactly from Eq.~(\ref{eq:3}) without any approximation and therefore, Eq.~(\ref{eq:5}) is irrelevant when $k\exp{[\lambda\xi]} = e^2/2\pi\epsilon_0R^3$. However, the proper question here is that whether Eq.~(\ref{eq:5}) when summed after taking $k\exp{[\lambda\xi]} = e^2/2\pi\epsilon_0R^3$ (not $(e^2/2\pi\epsilon_0R^3)^{\texttt{n}}\tilde{k}^{-\texttt{n}} \approx 0$ for $\texttt{n} \geq 4$) gives Eq.~(\ref{eq:8}). To see that it does, we rewrite Eq.~(\ref{eq:5}) accordingly 
\begin {eqnarray}
\lim_{\texttt{p} \rightarrow \infty}\tilde{E}(\xi) &=& \frac{\hbar\tilde{\omega}_0}{2}\bigg[\bigg(1 - \frac{1}{2} - \frac{1}{8} - \cdots\bigg) + \bigg(1 + \frac{1}{2} - \frac{1}{8} + \cdots \bigg)\bigg] - \hbar\tilde{\omega}_0 \nonumber \\&=& \bigg[\frac{\hbar\tilde{\omega}_0}{2}\sum_{\texttt{p} = 0}^{\infty}\frac{2\cdot(2\texttt{p})!}{(1 - 2\texttt{p})(\texttt{p}!)^2(4^{\texttt{p}})} - \hbar\tilde{\omega}_0\bigg] = \hbar\tilde{\omega}_0\bigg[\frac{\sqrt{2}}{2} - 1\bigg], \label{eq:6a}
\end {eqnarray}  
as it should be. Here $\texttt{p} \geq 0$, $\texttt{p} \in \mathbb{N}_{\rm even}$ and we have used $\sqrt{1+x} = \sum_{\texttt{p} = 0}^{\infty}\frac{(-1)^\texttt{p}(2\texttt{p})!}{(1 - 2\texttt{p})(\texttt{p}!)^2(4^{\texttt{p}})}x^{\texttt{p}}$ obtained from Ref.~\cite{abram}. Equation~(\ref{eq:6a}) unequivocally proves that the original vdW formula is for atoms with small polarization such that these atoms are weakly interacting. 

\section*{3. Results and Discussions}

\subsubsection*{{\rm 3.1}~\textit{Postmortem: stronger vdW attraction}}

{\color{blue} The interaction Hamiltonian, $H_{\rm I}$ in Eq.~(\ref{eq:2}) actually balances the repulsion and the attraction between two atoms if the system is in equilibrium. In other words, the polarization-induced standard vdW attraction is being balanced by the polarized electron-electron repulsion in accordance with Eq.~(\ref{eq:6}) (see Fig.~\ref{fig:2}(A)). Even though Eq.~(\ref{eq:6}) states that the atoms are attractive, but this is only true for certain range of $R$, $R_{\rm min} < R < R_{\rm max}$. This means that, for $R < R_{\rm min}$, the polarized electron-electron repulsion is activated to balance and to counter the attraction. In this case, a stronger attraction (stronger than Eq.~(\ref{eq:6})) is needed to overcome this repulsion. Using the new condition, $k\exp{[\lambda\xi]} = e^2/2\pi\epsilon_0R^3$ we derived Eq.~(\ref{eq:8}), which allows stronger attraction (because we have summed all the terms as proven in Eq.~(\ref{eq:6a})), and this stronger attraction is also balanced by the polarized electron-electron Coulomb repulsion (see Fig.~\ref{fig:2}(B)) for $R' < R'_{\rm min}$. In this second case, the electrons are more strongly polarized compared to Eq.~(\ref{eq:6}). These two cases mean that, for identical atoms, both Eqs.~(\ref{eq:6}) and~(\ref{eq:8}) cannot lead to an effective attraction due to polarized electron-electron repulsion, and therefore, they cannot initiate chemical reaction without external support, namely, collisions. Note this, these attractions exist because the Hamiltonian defined in Eq.~(\ref{eq:1new}) ignores the balancing force for $R_{\rm min} < R < R_{\rm max}$ that come from the polarized electron-electron repulsion, which are activated when and only when $R < R_{\rm min}$.} 

{\color{blue} Anyway, if the two atoms are not identical in their polarizability, then a third case can be constructed (see Fig.~\ref{fig:2}(C)) to unbalance the system and to understand why and how Eq.~(\ref{eq:8}) or equivalently, Eqs.~(\ref{eq:1}) and~(\ref{eq:1.2}) are physically handicapped due to the existence of polarized electron-electron repulsion (see the text before and after Eq.~(\ref{eq:1.3})). Therefore, we write down the interaction potential in the form given in Eq.~(\ref{eq:1.3}) that can give rise to an effective attraction such that the electron-electron repulsion is no longer enough to counter this attraction. In this final case, one is guaranteed to have an effective attraction between these two atoms because an additional electron-ion potential interaction term has been invoked to unbalance the two-atom system. We have to incorporate the above additional electron-ion interaction because Eq.~(\ref{eq:8}) cannot lead to an effective attraction due to polarized electron-electron repulsion. We will further elaborate on these three cases (Fig.~\ref{fig:2}(A,B,C)) in the following paragraphs.}

In fact, Eqs.~(\ref{eq:6}) and~(\ref{eq:8}) are true for any distance between two nuclei ($R$), for as long as there is no wavefunction overlapping. The difference here is that Eq.~(\ref{eq:6}) is valid only for systems with small polarization (or large isotropic polarization) and/or for weakly interacting atoms. Whereas, Eq.~(\ref{eq:8}) is for systems with large anisotropic polarization and/or for strongly interacting atoms. This means that, by definition, Eq.~(\ref{eq:6}) is not applicable for strongly interacting systems, while Eq.~(\ref{eq:8}) is inapplicable for weakly interacting systems. For example, for weakly interacting atoms (with small polarization or large isotropic polarization), we need to invoke the condition, $k\exp{[\lambda\xi]} \gg e^2/2\pi\epsilon_0R^3$ into Eq.~(\ref{eq:3}) to obtain Eq.~(\ref{eq:6}). On the other hand, we should impose $k\exp{[\lambda\xi]} = e^2/2\pi\epsilon_0R^3$ into Eq.~(\ref{eq:3}) to obtain Eq.~(\ref{eq:8}) for strongly interacting atoms (giving rise to large anisotropic polarization). Another point to note here is that Eq.~(\ref{eq:6}) can be regarded as valid for large $R$, while Eq.~(\ref{eq:8}) becomes valid for small $R$, which will be explained in the Section Further Analyses II.    

We now move on and rearrange Eq.~(\ref{eq:8}) to arrive at
\begin {eqnarray}
V_{\rm Waals}(R) = \bigg[\frac{\hbar^2e^2}{2m\pi\epsilon_0}\bigg]^{\frac{1}{2}}\bigg(\frac{1}{\sqrt{2}} - 1\bigg)\frac{1}{R^{3/2}}, \label{eq:1}
\end {eqnarray}  
\begin {eqnarray}
\tilde{V}_{\rm Waals}(\xi) = \hbar\omega_0\bigg(\frac{1}{\sqrt{2}} - 1\bigg)\exp{\bigg[\frac{1}{2}\lambda\xi\bigg]}, \label{eq:1.2}
\end {eqnarray}  
by using $\tilde{k} = e^2/2\pi\epsilon_0R^3$ interchangeably. Here, we should not assume that $\xi$ or $\tilde{k}$ has become $R$-dependent from the condition, $\tilde{k} = e^2/2\pi\epsilon_0R^3$. In fact, this condition simply means that the magnitude of $\tilde{k}$ is equal to the magnitude of $e^2/2\pi\epsilon_0R^3$. Both $\tilde{k}$ and $e^2/2\pi\epsilon_0R^3$ vary independently.  

Obviously, $V_{\rm Waals}(R)$ and $\tilde{V}_{\rm Waals}(\xi)$ are negative guaranteed by the inequality $1/\sqrt{2} - 1 < 0$, hence these interactions are always attractive, while the proportionality, $V_{\rm Waals}(R) \propto 1/R^{3/2}$ implies it is of a stronger-type compared to the renormalized standard vdW interaction, $\tilde{V}^{\rm std}_{\rm Waals}(\xi) \propto 1/R^{6}$ (see Fig.~\ref{fig:1}). It is worth noting that the proportionality $\tilde{V}_{\rm Waals} \propto 1/R^{3/2}$ is identical to the vdW interaction energies derived for two different geometric conditions~\cite{vili,biobook}. In particular, the relevant geometric conditions are (a) two infinite-length cylinders with radii $r_1$ and $r_2$, separated by $R$, and (b) a cylinder (radius $r_1$ and with infinite length) separated from a two-dimensional sheet (negligible thickness but with infinite length and width) at a distance $R$. Contrary to Eq.~(\ref{eq:1}), both (a) and (b) must satisfy the condition, $R \ll r_{1,2}$, in which, these classical objects are very close to each other~\cite{biobook}. In contrast, Eqs.~(\ref{eq:1}) and~(\ref{eq:1.2}) are for two spherical quantum objects (see Fig.~\ref{fig:1}) that satisfy conditions, $R \gg r_{1,2}$ and $k\exp{[\lambda\xi]} = e^2/2\pi\epsilon_0R^3$ (large polarization or weakly bound valence electrons). This large polarization will give rise to an anisotropic polarization in the presence of strong interactions between atoms. 

However, the imposed condition $R \gg r_{1,2}$ does not in any way, implies Eq.~(\ref{eq:1.2}) is inapplicable for any $R$ other than $R \gg r_{1,2}$. For example, what Eq.~(\ref{eq:1.2}) tells us is that for a given $R$ such that $R \gg r_{1,2}$, if $\xi$ is large, then one obtains a large attractive interaction. Now, the only way to change the value of $\xi$ is to change the type of atoms because each atom has a unique set of ionization energies or energy level spacings. Therefore, when we extrapolate this scenario for any $R$ such that $R \geq r_1 + r_2$, then such an extrapolation does not technically and physically change the effect of the renormalizer or $\xi$. In other words, large $\xi$ still gives rise to a large attractive vdW interaction, regardless whether the atoms are close to each other or far apart. 

Unfortunately, Eq.~(\ref{eq:1}) does not provide any new physical insight other than, that it is stronger than the original vdW interaction. Whereas, Eq.~(\ref{eq:1.2}) is indeed a function of the microscopic variable, $\xi$, but physically not yet appropriate because Eq.~(\ref{eq:1.2}) does not allow large anisotropic polarization effect--- the repulsive interaction between the anisotropically polarized electrons from two identical atoms separated by $R$, will give rise to an effective repulsion, not an attractive interaction (see Fig.~\ref{fig:2}(B)). As a consequence, Eq.~(\ref{eq:1.2}) certainly cannot lead to a stronger vdW attraction for large anisotropic polarization, due to the $e$-$e$ repulsion between anisotropically polarized atoms. Warning: Eq.~(\ref{eq:6}) is valid when $k\exp{[\lambda\xi]} \gg e^2/2\pi\epsilon_0R^3$ (valid for the least polarizable or weakly interacting atoms), whereas Eq.~(\ref{eq:1.2}) is valid for $k\exp{[\lambda\xi]} = e^2/2\pi\epsilon_0R^3$ (valid for easily polarizable or strongly interacting atoms). For example, the condition, $k\exp{[\lambda\xi]} \gg e^2/2\pi\epsilon_0R^3$, which allows us to invoke the solution in series (see Eq.~(\ref{eq:5})) imposes the least polarizability condition into the atomic system. In contrast, $k\exp{[\lambda\xi]} = e^2/2\pi\epsilon_0R^3$ by definition allows large polarization. We stress here that both Eqs.~(\ref{eq:6}) and~(\ref{eq:1.2}) ignore the existence of $e$-$e$ Coulomb repulsion, $V^{\rm e-e}_{\rm Coulomb}$ between polarized electrons.

Now, the easily polarizable atoms (that obey Eq.~(\ref{eq:1.2})) satisfies the anisotropic polarization (see Fig.~\ref{fig:2}(B)). Here, the polarized electrons will also activate the repulsive $e$-$e$ Coulomb interaction ($V^{\rm e-e}_{\rm Coulomb}$) between the polarized atoms. Therefore, we require $V^{\rm e-e}_{\rm Coulomb} + \tilde{V}_{\rm Waals}(\xi) < 0$ for an overall strong attraction. In fact, the existence and importance of $V^{\rm e-e}_{\rm Coulomb}$ between any two highly anisotropically polarizable entities has been analytically proven and discussed in Refs.~\cite{pccp,pla3,cpc}. However, the effect of anisotropic polarization, induced by strongly interacting atoms, which gives rise to the $e$-$e$ repulsive interaction ($V^{\rm e-e}_{\rm Coulomb}$) has been neglected in the standard (for both renormalized and original) vdW formula, given in Eq.~(\ref{eq:6}). 

Consequently, Eq.~(\ref{eq:1}), which allows stronger vdW attractive interaction needs to be made compatible with Eq.~(\ref{eq:1.2}) in the presence of large anisotropic polarization. In order to do so, we need to do something to counter the repulsive effect coming from $V^{\rm e-e}_{\rm Coulomb}$. We know Eq.~(\ref{eq:1.2}) allows large polarization but requires both atoms to be least polarizable (to avoid $V^{\rm e-e}_{\rm Coulomb}$) such that the nucleus of atom 1 (left-hand side (l.h.s) of Fig.~\ref{fig:1}) attracts the valence electron of atom 2 (right-hand side (r.h.s) of Fig.~\ref{fig:1}), and vice versa. Thus, Eq.~(\ref{eq:1.2}), which ignores $V^{\rm e-e}_{\rm Coulomb}$, cannot lead to a stronger vdW attraction. However, the strength of this attraction can be greatly enhanced if the polarizability of these two atoms are made to be strongly asymmetric (see Fig.~\ref{fig:2}(C)), \textit{i.e.}, these atoms are strongly interacting. Meaning, one atom is anisotropically polarized in the presence of another. In particular, if one of the atoms has large $\xi$, while the other has small $\xi$ ($\xi_{\rm l.h.s} < \xi_{\rm r.h.s}$) then this will guarantee the existence of $V^{\rm e-ion}_{\rm Coulomb} + \tilde{V}_{\rm Waals}(\xi) < 0$ because $V^{\rm e-ion}_{\rm Coulomb}$ is also negative by definition. In this case, the electron-ion (e-ion) Coulomb attraction ($V^{\rm e-ion}_{\rm Coulomb}$) is between the polarized valence electron of atom 1 (easily polarizable) and the nucleus of atom 2 (difficult to be polarized). 

To justify the stronger vdW attraction given in Eq.~(\ref{eq:1}), we require asymmetric polarizability between two atoms with strong interactions (see Fig.~\ref{fig:2}(C)). In this case, an asymmetric-polarization can be properly introduced into Eq.~(\ref{eq:1.2}) to obtain
\begin {eqnarray}
\tilde{V}'_{\rm Ramachandran}(\xi) = V^{\rm e-ion}_{\rm Coulomb} + \frac{1}{2}\tilde{V}_{\rm Waals}(\xi). \label{eq:1.3}
\end {eqnarray}  
Equation~(\ref{eq:1.3}) is the one that correctly corresponds to Eq.~(\ref{eq:1}) in the presence of large asymmetric polarization (see Fig.~\ref{fig:2}(C)). Our notation, $\tilde{V}'_{\rm Ramachandran}(\xi)$ is justified in the Section, Additional Notes (see the last paragraph). The second term on the r.h.s of Eq.~(\ref{eq:1.3}) represents the effect of two identical and weakly interacting atoms coming face-to-face (see Fig.~\ref{fig:1}), hence the factor $1/2$. Whereas, $V^{\rm e-ion}_{\rm Coulomb}$ captures the Coulomb attraction between a polarized electron from a polarizable atom (induced by the non-polarizable atom) and the nucleus of a non-polarizable atom. The screening effect is included by default whenever we deal with $\xi$ (due to Eq.~(\ref{eq:IN14b})) to predict the polarizability~\cite{cpc} (or any physical properties) from the ionization energy theory. 

In any case, the atoms depicted in Fig.~\ref{fig:1} are least polarizable (weakly interacting), and they affect each other equally with approximately isotropic polarization. While, $V^{\rm e-ion}_{\rm Coulomb}$ is defined as the interaction between the polarized electron of an easily-polarizable atom and the nucleus of the least polarizable nearest-neighbor atom. Therefore, we do not need to include the factor $1/2$ in $V^{\rm e-ion}_{\rm Coulomb}$ due to its definition given above, in which the induced anisotropic polarization is unidirectional, from the polarized valence electron of a strongly polarizable atom, to the nucleus of the least polarizable atom. In the absence of renormalization, the new condition reads $k = e^2/2\pi\epsilon_0R^3$, and one obtains an exact copy of Eq.~(\ref{eq:1}) for $V_{\rm Waals}(R)$ and
\begin {eqnarray}
V_{\rm Waals}(k) = \hbar\sqrt{\frac{k}{m}}\bigg(\frac{1}{\sqrt{2}} - 1\bigg). \label{eq:10}
\end {eqnarray}  
Here, Eq.~(\ref{eq:10}) physically correctly captures the same effect as Eq.~(\ref{eq:1.2}). In particular, for a given separation $R$, large $k$ implies small polarization, which is exactly identical to large $\xi$ that causes small polarization. Hence, two least polarizable atoms (weakly interacting) can give rise to the standard vdW attractive interaction, for instance, when $k \rightarrow$ maximum (from Eq.~(\ref{eq:10})), or $\xi \rightarrow$ maximum (from Eq.~(\ref{eq:1.2})) then both $V_{\rm Waals}(k)$ and $\tilde{V}_{\rm Waals}(\xi)$ increase to a maximum value due to zilch Coulomb repulsion between the polarized electrons (isotropically polarized atoms). As stated earlier, Eq.~(\ref{eq:1}) is independent of any microscopic variable (that can be related to the atomic structure), and as such it is not useful, except it correctly predicts the inverse proportionality between $V_{\rm Waals}$ and $R^{3/2}$. On the other hand, the changes in $k$ (from Eq.~(\ref{eq:10})) is arbitrary, for example, we will never know how to implement the changes in $k$ accordingly for different atoms in a given system. In contrast, and as proven earlier, only Eqs.~(\ref{eq:1.2}) and~(\ref{eq:1.3}) give us the option to accurately extract the microscopic information on two competing interactions, between the vdW attraction and the $e$-$e$ Coulomb repulsion (due to anisotropically polarized atoms) simultaneously, for strongly interacting atoms (\textit{via} $\xi$). 

\subsubsection*{{\rm 3.2}~\textit{Further analyses I: second-order perturbation theory}}

Equation~(\ref{eq:1}) is shown to vary as $1/R^{3/2}$, and this stronger vdW attraction can only be understood by adding an attractive Coulomb interaction (due to asymmetric polarization between two strongly interacting atoms) to a standard vdW attraction (Eq.~(\ref{eq:6})), or to a stronger vdW attraction (Eq.~(\ref{eq:1.2})). Therefore, due to this added Coulomb attraction, one cannot obtain Eq.~(\ref{eq:1}) using the 1/$R$ expansion since each individual energy term in this expansion represents a specific interaction (Coulomb, dipole, quadrupole, and so on) without any additional interaction (that gives rise to anisotropic polarization) added to 1/$R$ expansion~\cite{tai}. Meaning, each specific interaction is determined by a specific $\texttt{m}$ and varies as 1/$R^{\texttt{m}}$ where $\texttt{m} = 1, 2, \cdots$, which is applicable for weakly interacting systems. Recall here that a weakly interacting system means there will be no induced anisotropic polarization, even though these atoms can have large isotropic polarization due to intense laser. To understand this point, we derive Eq.~(\ref{eq:6}) using the second-order perturbation theory (the first-order energy is zero)~\cite{griffiths}. The renormalized second-order energy 
\begin {eqnarray}
&&\tilde{E}^{(2)}_{s} = \bigg|-\frac{e^2}{2\pi\epsilon_0R^3}\bigg|^2\sum_{s \neq t}^{\infty}\frac{|\langle\varphi^{(0)}_{t}(1)|r_1|\varphi^{(0)}_{s}(1)\rangle|^2|\langle\varphi^{(0)}_{t}(2)|r_2|\varphi^{(0)}_{s}(2)\rangle|^2}{\big[\tilde{E}^{(0)}_{s}(1) - \tilde{E}^{(0)}_{t}(1)\big] + \big[\tilde{E}^{(0)}_{s}(2) - \tilde{E}^{(0)}_{t}(2)\big]}, \label{eq:11}
\end {eqnarray}  
where (1) and (2) refer to atom 1 and 2 (see Fig.~\ref{fig:1}), while the subscripts, $s$ and $t$ denote the ground and excited eigenvalues ($\tilde{E}_{s,t}$) and eigenstates ($\varphi_{s,t}$), respectively. Moreover, 
\begin {eqnarray}
H_{\rm I} \approx -\frac{e^2r_1r_2}{2\pi\epsilon_0R^3}, \label{eq:11x}
\end {eqnarray}
after taking $R \gg r_{1,2}$. We first note 
\begin {eqnarray}
&&\tilde{E}^{(0)}_{s}(1) - \tilde{E}^{(0)}_{t}(1) + \tilde{E}^{(0)}_{s}(2) - \tilde{E}^{(0)}_{t}(2) = 2\bigg[\frac{1}{2}\hbar\tilde{\omega}_0 - \frac{3}{2}\hbar\tilde{\omega}_0\bigg], \label{eq:12}
\end {eqnarray}  
where $(1/2)\hbar\tilde{\omega}_0 > 0$ is the ground-state eigenvalue, while $(3/2)\hbar\tilde{\omega}_0 > 0$ is the first excited state eigenvalue for the harmonic oscillator-like two atomic system sketched in Fig.~\ref{fig:1}. Subsequently, using Eq.~(\ref{eq:12}) and
\begin {eqnarray}
&&|\langle\varphi^{(0)}_{t}(1)|r_1|\varphi^{(0)}_{s}(1)\rangle|^2|\langle\varphi^{(0)}_{t}(2)|r_2|\varphi^{(0)}_{s}(2)\rangle|^2 = \bigg[\frac{\hbar}{2m\tilde{\omega}_0}\bigg]^2, \label{eq:13}
\end {eqnarray}  
we can derive Eq.~(\ref{eq:6}) from Eq.~(\ref{eq:11}) following Griffiths~\cite{griffiths}--- the renormalization procedure introduced here does not disturb the derivation in any way. As predicted earlier, $\tilde{E}^{(2)}_{s} \propto 1/R^{6}$, and is always negative determined entirely by the eigenvalue-denominator (ground state eigenvalue $-$ excited state eigenvalue = $-2\hbar\tilde{\omega}_0$) in Eq.~(\ref{eq:12}). The numerator in Eq.~(\ref{eq:11}) is positive-definite. Interestingly, the same conclusions (without renormalization) can be obtained from Ref.~\cite{tai}. For example, the coefficient $C_6$ [see Eq.~(4.3) in Ref.~\cite{tai}] represents the vdW attraction, and is always negative due to $\epsilon_{a}(k_a) - \epsilon_{a}(k_a') + \epsilon_{b}(k_b) - \epsilon_{b}(k_b') < 0$ because $\epsilon_{a,b}(k_{a,b},k_{a,b}') < 0$, $|\epsilon_{a}(k_a')| < |\epsilon_{a}(k_a)|$ and $|\epsilon_{b}(k_b')| < |\epsilon_{b}(k_b)|$. Here, $a$ and $b$ denote atom 1 and 2, respectively, $k_{a,b}$ and $k_{a,b}'$ represent both the principal and the angular momentum quantum numbers such that $k_{a,b} \neq k_{a,b}'$, while $\epsilon_{a,b}$ are the respective eigenvalues. 

To understand the above inequalities, assume $a$ and $b$ are two hydrogen atoms, and therefore, $\epsilon_{a}(k_a)$ and $\epsilon_{b}(k_b)$ are the ground-state eigenvalues for these atomic hydrogen, respectively, while $\epsilon_{a}(k_a')$ and $\epsilon_{b}(k_b')$ are the respective excited-state eigenvalues for hydrogen atoms. Since, $E^{\rm ground}_1 = -$13.6 eV $= \epsilon_{a}(k_a) = \epsilon_{b}(k_b)$, $E^{\rm excited}_2 = -$3.4 eV $= \epsilon_{a}(k_a') = \epsilon_{b}(k_b')$, then we can immediately see why the denominator [in Eq.~(4.6) of Ref.~\cite{tai}] is always negative (as predicted earlier) for as long as the polarization is from a lower energy level (strongly bounded electron) to a higher energy level (weakly bounded electron).   

Add to that, $C_6 \propto 1/R^{6}$ originates from the interaction potential operator, $V_3 \propto 1/R^{3}$ [see Eqs.~(2.13) and~(4.3) in Ref.~\cite{tai}]. In our atomic system, we had $H_{\rm I} \propto -1/R^{3}$ [see Eq.~(\ref{eq:11x})]. Importantly, regardless whether $V_3$ is negative or positive, $C_6$ is guaranteed to be negative due to the eigenvalue-inequalities given above, and note that $|\langle\cdots|\pm V_3|\cdots\rangle|^2 \propto 1/R^6$, which is positive-definite. As anticipated, $V_3$ is a specific interaction potential ($\texttt{m} = 3$), which was identified to give the vdW attraction without any additional induced anisotropic polarization added to $V_3$. Therefore, similar to Eq.~(\ref{eq:6}), $V_3$ excludes the $e$-$e$ Coulomb repulsion in the presence of large anisotropic polarization, both in the ground and excited states or in the mixed states. 

\subsubsection*{{\rm 3.3}~\textit{Further analyses II: multipole expansion}}

The large $e$-$e$ repulsion due to large anisotropic polarization (see Fig.~\ref{fig:2}(B)) has been excluded by invoking certain initial conditions used in Ref.~\cite{tai} to write the potential, 
\begin {eqnarray}  
V = \sum_{\texttt{m}=3}^{\infty}\frac{V_{\texttt{m}}}{R^{\texttt{m}}} \label{eq:14}. 
\end {eqnarray}  
The initial conditions were $R > r_a + r_b$ such that most of the charges are distributed within an atomic or a molecular radius. This means that the allowable polarization is either small or increase isotropically (see Fig.~\ref{fig:2}(A)). Isotropically increasing polarization implies that the amount of increased polarization is essentially identical in all physical directions, namely, in $x$, $y$ and $z$ axes. Now, these conditions need to be imposed so that Eq.~(2.14) in Ref.~\cite{tai} is valid. Rightly so, Chang has correctly invoked the above conditions even for large atomic or molecular polarization such that the amount of increased polarization is three-dimensionally isotropic. This in turn means that the systems studied there were for weakly interacting atoms or molecules. If the atoms or molecules are strongly interacting, then any polarization that increased isotropically has got to be invalid. In addition, the condition that requires the polarization to increase isotropically cannot hold for $R \approx r_a + r_b$. 

Hence, what we have proven earlier is that isotropic polarization is inapplicable for $R \approx r_a + r_b$ (for small internuclear distances) because we need to take the anisotropic polarization effect into account, which has been included in our formalism [see the first term in Eq.~(\ref{eq:1.3})]. Moreover, one should also note that the isotropic polarization is not applicable for strongly interacting systems, even if $R > r_a + r_b$. Thus, Eq.~(\ref{eq:1.3}) is derived for strongly interacting systems that allow asymmetric polarization for $R > r_a + r_b$ (see Fig.~\ref{fig:2}(C)). In contrast, isotropic polarization can only occur for systems that are weakly interacting and $R > r_a + r_b$. It is physically incorrect to enforce isotropic polarization for $R \approx r_a + r_b$ or for strongly interacting systems. In particular, for strongly interacting systems with $R \geq r_a + r_b$, there will definitely be a significant amount of induced anisotropic polarization on one atom or molecule that is caused by its neighboring atoms or molecules, which then gives rise to this strong asymmetric polarization. This is what we have addressed in this paper when we derived Eq.~(\ref{eq:1.3}). One may correctly wonder if there is an effective $e$-$e$ Coulomb repulsion between two isotropically polarized atoms or molecules. The effective repulsion will only be activated when the electrons in both atoms or molecules are polarized anisotropically in the presence of strong interaction between these atoms or molecules, and/or when $R \approx r_a + r_b$. For example, atoms or molecules will be isotropically polarized in the gas-phase, if these atoms or molecules are weakly interacting in the presence of high temperature or intense laser. This isotropic polarization can be large. In this case, Eqs.~(\ref{eq:2}) and~(\ref{eq:3}) are applicable where one just starts with a new set of radii for atom 1 and 2, and again obtains Eq.~(\ref{eq:6}). The new set of radii are larger than the original radii because the electrons are now polarized isotropically (see Fig.~\ref{fig:2}(A)). In fact, one can go on and improve Eq.~(\ref{eq:6}) by incorporating other leading terms due to spin-orbit coupling and quadrupole-quadrupole interaction as carried out by Chang~\cite{tai}. But this additional effects cannot induce the $e$-$e$ Coulomb repulsion, which is due to anisotropic polarization, which occurs exclusively when there is a strong interaction between atoms or molecules or when the average internuclear distance, $R \approx r_a + r_b$ (that eventually will induce the strong interaction). 

Now, we should know that the standard vdW theory cannot handle anisotropically polarized atoms or molecules even though it has incorporated other leading terms, namely, $1/R^5$ (quadrupole-quadrupole) and $1/R^3$ (magnetic interaction or due to spin-orbit coupling). These terms do improve the potential given in Eq.~(\ref{eq:14}) by incorporating the above stated physical mechanisms, and can be used to improve Eq.~(\ref{eq:11x}), but they (the other leading terms) do not cause the $e$-$e$ Coulomb repulsion to exist. For example, these other leading terms are there to improve the equation for $C_6$ or Eq.~(\ref{eq:6}) by incorporating the appropriate correction terms such that each term represents a specific physical effect, namely, $1/R^3$ represents magnetic interaction, whereas $1/R^5$ is due to quadrupole-quadrupole interaction. In this case, the Coulomb repulsion is not a correction term, but originates from the $e$-$e$ repulsion due to polarized electrons. This means that, the first term in Eq.~(\ref{eq:1.3}) is the leading term that represents the asymmetric polarization, and is responsible for the stronger vdW attraction in strongly interacting systems. In particular, the above $e$-$e$ Coulomb repulsion, which is due to the large anisotropic polarization, caused by the strongly interacting atoms or molecules, can only be overcome by the first term in Eq.~(\ref{eq:1.3}). As a matter of fact, Eq.~(2.14) in Ref.~\cite{tai} has been enforced to make sure that the standard vdW theory presented in Ref.~\cite{tai} is for isotropic polarizations, which does not take large anisotropic polarization effect into account. For example, see the spherical harmonics given on page 139 in Ref.~\cite{griffiths} and Eq.~(2.15) in Ref.~\cite{tai} to observe the expansion for $V_{\texttt{m}}$ that is isotropic by definition, which has been enforced in Ref.~\cite{tai}. For example, the standard vdW theory given in Ref.~\cite{tai} is valid for weakly interacting system and for $R > r_a + r_b$ such that the polarizations are (assumed to be) approximately isotropic. On the other hand, the vdW theory with asymmetric polarization developed here is valid for strongly interacting systems, and for $R \geq r_a + r_b$. 

In summary, a new van der Waals (vdW) interaction of a stronger type is proven to exist analytically by generalizing the original vdW interaction. This generalization includes three types of polarizations between two atoms, namely, isotropic, anisotropic and asymmetric polarizations. We show that an isotropic polarization is only valid for $R \gg 2r$ ($r = r_a = r_b$), which is a well known result within the original vdW theory. But the other two polarizations (anisotropic and asymmetric) are shown to be valid for all $R$. Recall here that $R$ is the internuclear distance between two atoms, while $r$ is the atomic radius for two identical atoms. Using the above generalization, we went on to explain why and how the original vdW attraction does not lead to an attractive interaction between two highly polarizable atoms due to large Coulomb repulsion between polarized electrons. This explains why $R \gg r$ is a necessary condition for systems with isotropic polarizations, in accordance with the original vdW formalism. However, for $R \approx 2r$, the two atomic system should be considered as strongly interacting, giving rise to an anisotropic or an asymmetric polarization. In these cases, there is no requirement on $R$ such that these polarizations are valid for any $R$.

\section*{4. Additional notes}

Here, we briefly reiterate the consequences of our main result, which are important for an overall understanding and also to avoid confusion within quantum chemistry. If one employs a full electronic Hamiltonian (within the quantum chemical methods) for different interaction strengths, then it is indeed possible to calculate the reaction barriers, and the reaction rates accurately by means of the variational principle and approximate wavefunctions~\cite{selva,senthil,niu,kummer}. On the other hand, the theory presented here has been used to predict (exactly) which atoms (from two molecules) can react to form a peptide bond, in accordance with the experimental observations using the principle of maximum interaction~\cite{jcs1}. Here, we did not attempt to calculate the activation energies because our methodology, by definition, does not allow such calculations to be carried out because our theory does not have variationally adjustable parameters and does not rely on wavefunctions directly. Therefore, our motivation here is to discover the precise microscopic physical origin responsible for a chemical reaction between two atoms, which is still unclear within the quantum chemical calculations due to the introduction of variationally adjustable interaction terms and wavefunctions.

Here, we have derived the relevant interactions (Coulomb and van der Waals types) that control a particular chemical reaction where such interactions cannot be obtained by renormalizing or adjusting the coefficient, $C_6$ (discussed earlier). For example, we obtained the stronger vdW interaction by summing all the terms in Eq.~(\ref{eq:6a}) (with or without renormalization), and by adding the electron(from atom 1)-ion(from atom 2) Coulomb attraction. In fact, $C_6 \propto |V_3|^2$ is a natural consequence of Eq.~(\ref{eq:2}) for $R \gg r_{1,2}$ (see Eq.~(\ref{eq:11x})), which then leads to the standard vdW term (if we only take the first three terms in Eq.~(\ref{eq:6a})). In this work, the stronger interaction (see Eq.~(\ref{eq:1.3})) is shown to go beyond adjusting or renormalizing Eq.~(\ref{eq:6a}), which has been used to understand precisely how two reactant atoms can overcome the activation energy via collisions (for weakly interacting atoms), or without it (for strongly interacting atoms with induced anisotropic polarization). In fact, we also have justified that the electron(from atom 1)-electron(from atom 2) Coulomb repulsion is at least, one of the main sources that defines the origin of activation energy.

We now realized that the discovery of stronger attraction between non-chemically bonded (or simply nonbonded) atoms has been reported much earlier by Ramachandran, Sasisekharan and Ramakrishnan~\cite{ramc,ramc2} where nonbonded atoms in amino acids (C, N, O and H) can approach much closer than what is allowed by the standard vdW bond. Ramachandran correctly and properly reasoned that~\cite{ramc,ramc2} the above nonbonded atoms in amino acids and peptides can approach closer (with shorter distance between nonbonded atoms) leading to the formation of triple-helical structure in collagen. He further proved that such a stronger attraction due to allowed bond rotation, did play an important role in polypeptide chain's conformations, limited only by the steric-hindrance, which led him, Sasisekharan and Ramakrishnan to invent the Ramachandran plot\cite{ramplot}. In view of this fact, we now set the record straight by stating that Eq.~(\ref{eq:1.3}) rediscovers the Ramachandran bond (in the absence of chemical reaction) such that the standard vdW, hydrogen, non-covalent carbon and other similar bonds~\cite{rath,biman,roy,reza} are special cases.
 
\section*{5. Conclusions}

Using the ionization energy theory and the energy-level spacing renormalization group method, we obtained the relevant analytic functions to formally expose the limitation in the original vdW formalism. In particular, the Coulomb repulsion between polarized electrons has been neglected by requiring small polarization, which has been implicitly embedded into the original formalism by requiring the interaction potential constant ($k$) to be much greater than the Coulomb interaction between atoms (molecules). Or, by imposing a large isotropic polarization by invoking spherical harmonics. After taking the large anisotropic polarization effect into account, we managed to obtain a new vdW formula of a stronger type, which is proportional to $1/R^{3/2}$. However, the new formalism also leads to a formula that is proportional to $\exp[\lambda{\xi}]$ where $\xi$ is the energy level spacing of a given molecule or system. Here, large $\xi$ means large $k$, in accordance with the original vdW theory. Therefore, large anisotropic polarization does not favor large attractive interaction at all, because of the $e$-$e$ Coulomb repulsion effect (due to polarized electrons). 

To overcome this intrinsic problem, large asymmetric polarization between two interacting molecules was invoked to expose the existence of a stronger vdW interaction, in which, the above repulsion has been physically converted into a Coulomb attraction \textit{via} an asymmetric polarization. For example, the attraction between the polarized electrons (of an easily anisotropically polarizable atom) and a nearest neighbor nucleus (of the least polarizable atom) gives rise to this asymmetric polarization. This means that, the large vdW interaction (due to $1/R^{3/2}$) can be justified with the large asymmetric polarization effect. Therefore, there exists a stronger attractive vdW interaction between two strongly interacting atoms if they are polarized asymmetrically such that one of the atom or molecule is anisotropically polarized due to interaction with its nearest least-polarizable neighbor. 

The overall strong attractions that come from the stronger ion-ion vdW and electron-ion Coulomb interactions determine the possibility of a chemical reaction between two or more neutral atoms, for a given set of conditions (chemical potential, temperature, pressure and concentration). In particular, we have established that (i) a chemical reaction proceeds as a result of some stronger attractions between reactant atoms, (ii) the attractions are due to stronger vdW and electron-ion Coulomb interactions, (iii) systems that contains a mixture of least and easily polarizable reactant atoms induce asymmetric polarization, which then leads to the stronger attractions, and (iv) these stronger attractions are the ones that eventually overcome the so-called activation energy. However, it turns out to be that this stronger attraction has been discovered and proven to exist much earlier by Ramachandran, Sasisekharan and Ramakrishnan. Rightly so, the bond caused by this generalized stronger attraction (see Eq.~(\ref{eq:1.3})) is (and should be) called the Ramachandran bond. We have shown here that this stronger attraction can then lead to chemical reaction, if the conditions stated above are met~\cite{jcs1}.

\section*{Acknowledgments}

I am grateful to Madam Sebastiammal Savarimuthu, Mr. Arulsamy Innasimuthu, Madam Amelia Das Anthony, Mr. Malcolm Anandraj and Mr. Kingston Kisshenraj for their financial support and kind hospitality between August 2011 and August 2013. I also would like to thank the referees for directing me to Ref.~\cite{tai}, and for pointing out all the missing explanation. Special thanks to Dr. Mir Massoud Aghili Yajadda (Commonwealth Scientific and Industrial Research Organization, Melbourne) and Mr. Alexander Jeffrey Hinde (The University of Sydney) for providing some of the listed references.

\end{document}